\documentclass[runningheads]{llncs}
\usepackage{graphicx}

\begin{document}
\title{Intelligence Education made in Europe}
\titlerunning{Intelligence Education made in Europe}
\author{Lars Berger \inst{1} 
\and Uwe M. Borghoff \inst{2}\thanks{Corresponding author: uwe.borghoff@unibw.de}
\and Gerhard Conrad \inst{1} 
\and Stefan Pickl \inst{2}
}
\authorrunning{L. Berger, U. M. Borghoff, G. Conrad \& S. Pickl}

\institute{Department of Intelligence \\ Federal University of Administrative Sciences, Berlin, Germany
\and
Department of Computer Science \\ University of the Bundeswehr Munich, Neubiberg, Germany \\
}
\maketitle              

\begin{abstract}
Global conflicts and trouble spots have thrown the world into turmoil. Intelligence services have never been as necessary as they are today when it comes to providing political decision-makers with concrete, accurate, and up-to-date decision-making knowledge. This requires a common co-operation, a common working language and a common understanding of each other. The best way to create this “intelligence community” is through a harmonized intelligence education.

In this paper, we show how joint intelligence education can succeed. We draw on the experience of Germany, where all intelligence services and the Bundeswehr are academically educated together in a single degree program that lays the foundations for a common working language. We also show how these experiences have been successfully transferred to a European level, namely to ICE, the Intelligence College in Europe. Our experience has shown that three aspects are particularly important: firstly, interdisciplinarity or better, transdisciplinarity, secondly, the integration of IT knowhow and thirdly, the development and learning of methodological skills. Using the example of the cyber intelligence module with a special focus on data-driven decision support, additionally with its many points of reference to numerous other academic modules, we show how the specific analytic methodology presented is embedded in our specific European teaching context.

\keywords{Intelligence College in Europe (ICE) 
\and cyber intelligence
\and intelligence community
\and transdisciplinary education 
\and IT knowhow 
\and methodological skills
\and Intelligence Service Oriented Approach (ISOA)}
\end{abstract}

\section{Introduction}
Current intelligence discourse increasingly emphasizes the need for an interdisciplinary approach to address the diverse challenges of the 21st century security landscape. This recognition underscores the intricate interplay of diverse disciplines such as history, political science, sociology, psychology, computer science, and law to fully understand and effectively respond to emerging threats. Intelligence research and practice require interaction and mutual understanding of different perspectives and methodologies to provide sound insights and strategic solutions in the future \cite{Fingar2023}.

In addition, rapid advances in technology have changed the nature of security threats, resulting in hybrid threats that exploit vulnerabilities in multiple domains. The collision of new technologies such as artificial intelligence (AI), cyber capabilities \cite{bonfanti2018cyber}, and biotechnology with human ingenuity poses unprecedented challenges to traditional security paradigms. As a result, intelligence professionals must constantly adapt and innovate to effectively counter these evolving threats. This ability to adapt and innovate requires sound interdisciplinary education and training, preferably over a lifetime. This includes not only technical skills, but also methodological skills, critical thinking, adaptability, and intercultural competence.

Intelligence education should also foster a culture of collaboration and information sharing that recognizes the interconnectedness of global security challenges \cite{glees2018intelligence}. By facilitating interdisciplinary dialogue, teaching and experiential learning opportunities \cite{burch2016towards}, educational institutions can cultivate a cadre of intelligence professionals capable of addressing the complexities of hybrid threats \cite{steingartner2021threat}. 

This is where the Intelligence College in Europe (ICE) comes 
in \cite{journals/corr/abs-2312.17107}. 
The main objective of ICE is to unite European national intelligence services for non-operational cooperation in order to overcome the current fragmentation of European intelligence services \cite{prunckun2016handbook}.  
ICE will celebrate its fifth anniversary in 2024 and has made significant progress in three key areas: First, as an ICE membership and participation initiative with 31 member countries, including 89 intelligence and security agencies. Second, as an academic network with 33 academic institutions in 18 member countries, ICE promotes academic exchange and research. Finally, ICE offers an academic program of more than 30 events, training and outreach activities for practitioners and 
academics alike.\footnote{https://www.intelligence-college-europe.org/} 

The academic program is based on two pillars. One pillar is the {\em postgraduate program}, which includes the {\em Cyber Intelligence} module presented in the following sections. Other modules offered here include {\em Counterterrorism} (hosted by Germany), {\em Intelligence and the Military} (hosted by the Netherlands), and {\em Societal Resilience to Hybrid Threats} (hosted by Romania with contributions from Croatia, Estonia, Finland, and Germany).
The other pillar is the {\em executive education program}. Both types of programs differ in the way they are run, in their content, and in the audiences they address among the intelligence professionals who have come together under the umbrella of the Intelligence College in Europe (ICE).
\begin{itemize}
\item
The {\em postgraduate program} usually features a smaller number of participants with a cap on numbers that ensures the feel and atmosphere of a postgraduate level seminar at a civilian university. The idea here is to offer an overview of where the cutting edge of relevant academic research currently stands in a way that is accessible and relevant to intelligence professionals at an earlier point in their careers. Teaching is entirely delivered by professors from academic institutions that make up the academic network within ICE. Participation in one module does not require participation in another module. Participating countries and intelligence services can thus send those professionals with the closest thematic fit.
\item
The {\em executive education program} by contrast is run with the intention that participants take part in every one of the up to half a dozen events of this category per academic year. The focus here is on intelligence practice with intelligence practitioners from hosting countries leading or contributing to most sessions in each event. The target audience is less early- or mid-career subject specialists, but those individuals that already take up senior positions in their services or expect to do so in the future.
\end{itemize}

For its academic program, ICE also draws on the experience of Germany's Master of Intelligence and Security Studies (MISS) \cite{books/sp/17/BorghoffD17}.  
To date, this is the only intelligence studies program in Germany that brings together all intelligence services and the Bundeswehr \cite{Scheffler2016}. 
MISS is also open to members of German ministerial administrations with links to security policy (in particular the Federal Chancellery, the Federal Ministry of the Interior, the Federal Ministry of Defense and the Federal Foreign Office) and members of parliamentary administrations involved in parliamentary control of the intelligence services.\footnote{https://www.unibw.de/ciss-en/miss} 

\section{Journey from Traditional National Intelligence Efforts \\ to a Concerted Common Approach through ICE: \\ The Role of Digitalization}
In the face of accelerating and widespread social and political upheavals caused by the effects of globalization, the ever-increasing digitalization of life \cite{mandras2022digital},  the effects of climate change, major demographic trends, technologies with unseen qualities of promise and threat, and the emergence of new centers of power \cite{ryan2022war},  instability, and conflict, intelligence organizations, as part of their societies and state structures, will need to adapt in order to survive and deliver \cite{gill2018intelligence}.  

Transformation will therefore be a core challenge for the services in the coming decades. It will have to be identified, understood and implemented by the services concerned and their oversight bodies. Wherever possible and appropriate, this process should be facilitated through consultation and exchange with like-minded services and institutions. This is where ICE and its specific module ``Cyber in its Implications for Intelligence, Analysis and Decision Making'' have their place and relevance \cite{fingar2011reducing}.  
Before presenting the new curriculum of this ICE module, it may be useful to discuss two more general questions:
\begin{itemize}
\item
What specific needs should such a pilot course address? 
\item
What are the leadership strategies for developing a comprehensive curriculum in which the pilot course is embedded?
\end{itemize}
To answer these fundamental questions, we must first characterize and analyze the specific function of intelligence organizations in political and military decision-making processes.

\subsection{Intelligence Organizations}
In their support role for political, security and defense decision-making, intelligence organizations are more relevant and important today than ever before in the face of an increasingly dynamic and multidimensional threat landscape. Adequate holistic situational awareness must be based on information superiority, which in turn must be ensured by the most advanced human and technical sensors for collection, as well as excellence in information management, assessment and reporting, which in turn must fit seamlessly into powerful, comprehensive decision making and implementation processes \cite{munir2022situational}.  

For all these dimensions, a conceptual framework for IT-based decision support is a key prerequisite. It addresses the following tasks: 
\medskip

\fbox{\centering%
    \parbox{.9\textwidth}{\bf\em%
        Collect and assess: \\ Data-driven processes will be the name of the game
    }%
}\medskip

Collect and assess relevant classified and publicly available information in core areas of national political, military, and security interest, such as early warning, current and emerging threats, adversary capabilities and intentions, emerging technologies and their implications, environmental risks and their impact on stability and prosperity, and migration drivers.
\medskip

\fbox{\centering%
    \parbox{.9\textwidth}{\bf\em%
        Comprehensive knowledge base and strategic intelligence \\ operations
    }%
}\medskip

Build an IT-driven, intelligence-led, comprehensive scientific knowledge base that has two core functions: First, to enable the Service to provide strategic intelligence to decision makers on developments, trends, risks, and threats emerging abroad that may affect the security interests of the country and its allies. Second, to provide comprehensive ad hoc background information and analysis in the event of imminent threats and strategic/tactical surprises. In a crisis, the closest thing to real-time information and awareness must be the gold standard for successful decision-making. 
\medskip

\fbox{\centering%
    \parbox{.9\textwidth}{\bf\em%
        World-class IT-based decision support and cyber capabilities
    }%
}\medskip

Build a resilient and powerful capability for debunking fake news in the face of the explosion of AI \cite{lin2020allies} and quantum-based disinformation and manipulation spreading \cite{qu2024qmfnd} through cyberspace in all its dimensions (data services, traditional media, social media, etc.) \cite{sharfman1995intelligence}.  This capability rests on two pillars: Information superiority based on the aforementioned knowledge base for debunking fake content itself, and high-level IT and cyber capabilities for detecting and analyzing the technical origin and background of fake content. 
\medskip

\fbox{\centering%
    \parbox{.9\textwidth}{\bf\em%
        Comprehensive approach: Operational and political levels
    }%
}\medskip

There is a need to maintain and continuously update a consolidated, nuanced, and comprehensive understanding of domestic and international trends, risks, or structures in society that affect national security in order to inform, educate, and empower decision-makers at the operational or political level, in the police or security services, and in government or parliament. There is an imperative need to maintain and improve strong counterintelligence and counterespionage capabilities in the face of a significant increase in the impact of hostile hybrid efforts by powerful international actors with ambitions for global or regional power projection and dominance. 
\medskip

\fbox{\centering%
    \parbox{.9\textwidth}{\bf\em%
        Optimizing counterintelligence and \\ counterterrorism capabilities
within a risk framework
    }%
}\medskip

These actors pose a risk to national security either directly or indirectly by harming vital national security interests abroad: By attacking their own citizens and facilities in third countries, by disrupting local infrastructure there needed to produce or deliver goods of value for domestic production or consumption, or by disrupting international or regional transportation or communications infrastructure, thereby directly affecting domestic needs. All of this is supported, enhanced, or even driven by major advances in communications and data-driven technologies \cite{smallwarsJ} that dramatically increase the speed of events, their global impact, and their specific ripple effects. 
\medskip

\fbox{\centering%
    \parbox{.9\textwidth}{\bf\em%
        Create, deliver and update situation awareness 
    }%
}\medskip

In terms of capability building, intelligence organizations must keep pace with, and where appropriate and possible, lead, these emerging megatrends of modern and future life. They will need to be technologically savvy and agile, but also highly competent in all matters of substance relevant to the security and vital interests of their governments, societies and states. Competence excellence must be the stated ambition and top agenda of intelligence organizations. They must be dynamic producers of world-class knowledge and situational awareness for decision-makers, and by no means a kind of self-sufficient static administration. 

All of this provides the framework and background for an international curriculum offered by ICE that focuses on the following question:

\begin{itemize}
\item
How to address current and future dynamic and comprehensive strategic threats and risks, and how to empower intelligence organizations to do so?
\end{itemize}
Of course, every country has its own geopolitical peculiarities. It has its own political and constitutional order and principles. It may have ``eternal interests'' and a specific set of risks and threats to mitigate. It may have specific structures and competencies to create and maintain situational awareness for decision-makers in the various spheres of governmental and societal activity and responsibility. Irrespective of these specific frameworks, the core requirement must be for the state and society to create high quality, resilient and capable capabilities that will enable them to keep pace with and meet the security challenges that lie ahead. Intelligence and security services themselves may have more limited mandates than those described above. However, these should then fit into a broader set of policy-making structures, best placed under the authority and coordinating power of a National Security Council, that contribute to that kind of vital comprehensive situational awareness in security-related matters, which in turn is a core foundation for state resilience and survival.

\subsection{Capability-Oriented Curriculum}
Based on a clear understanding of this future role of intelligence organizations in general, the following questions could serve as a baseline for a specific capability-oriented curriculum:
\begin{itemize}
\item
What professional skills are needed to fulfill the various missions?
\item
What kind of software and hardware will need to be available, taking into account dynamic developments in development and procurement?
\item
What kind of underlying technology will need to be implemented and evolved for the various tasks? 
\item
What kind of processes are needed to aggregate and provide structured databases that are accessible for meaningful data analysis? How will they be delivered? 
\item
What will be the quantitative and qualitative framework for hiring and managing people to staff and drive the structures? 
\end{itemize}

The innovative ICE curriculum is designed to address these issues in a holistic manner. The curriculum also relates skills to a specific mission-oriented approach that focuses on the following questions:
\begin{itemize}
\item
What exactly should the mission deliver, and in which fields of action? Strategic or tactical/up-to-the-minute intelligence? 
\item
How its work would be included in decision making processes of the respective customers? Operational support to political, military or police actions?
\item
Who are its customers and what are their needs and interests? 
\end{itemize}

In addition, the curriculum must be open to the diversity of government and service leadership environments: Some governments may have already implemented major transformational projects, others may not have done so or have done so only in a very limited way, in other cases the basic framework may already be outdated in some aspects. 

Because of its international and inter-service approach, the proposed curriculum will need to respect these different levels and perspectives, which each nation and service will address in its own way, without unwarranted outside interference:
\begin{itemize}
\item
What new technologies and capabilities including databases are available at the national level? To what extent and in what ways are AI \cite{NATOISR} and quantum computing \cite{TenThings} already being used in OSINT, SIGINT, CYBERINT, COMINT, SOCMINT, IMINT, GEOINT, storage, tools, collection \& exploitation, analysis, reporting, and decision making? What are the prospects for evolution?  
\item
At what level of professionalism is targeting and collection being conducted by the services? 
\item
Is there an overview of available or at least studied targets for HUMINT exploitation or technical exploitation and quantitative and/or qualitative gaps with respect to the mission and the adequacy of collection capabilities in terms of quantity and quality with respect to the mission? 
\item
What is the critical level of productivity, reach, and impact, taking into account timeliness, relevance of products, accessibility to customers, and usability in terms of customer horizons, needs, and working methods? 
\item
Is there an analysis of the national legal framework, taking into account legal mandates according to roles and responsibilities, and identifying gaps and shortcomings that affect service delivery? 
\end{itemize}

Despite the diversity of challenges and solutions at the respective national levels and frameworks, all nations are currently facing similar high-tech issues. This is the common ground upon which this cyber intelligence module can build and may be relevant to its participants within ICE \cite{warner2017intelligence}.  

\section{High Tech Matters: AI and Quantum Computing}
For all nations, AI and quantum computing will define and revolutionize all areas of intelligence collection, processing, analysis, production and distribution in the coming years/decades \cite{C4ISRNet}.   
Encryption must be made quantum-proof. Otherwise, services and governments risk losing their secrets and expertise to an adversary or to the general public first, and then to all sorts of malicious actors. That's a mortal danger for the services and everyone who relies on them. Quantum computing and its implications must therefore be a top priority for leadership in this area.

\subsection{Optimization of Decision Making Processes}
In addition, quantum computing has the groundbreaking potential to dramatically speed up the collection, analysis, communication, and decision-making processes. This applies not only to the services, but to the entire arena of political, military, and security decision-making processes, as well as the resulting actions themselves. Keeping pace with the dramatic acceleration of processes will therefore be a major challenge. Adequate structures and procedures will have to be developed and implemented - quickly - not only in the intelligence services, but also in the area of decision making \cite{Butler2021}.  Human decision-makers, with their inherently limited cognitive capabilities, must not be overwhelmed by the situational dynamics  \cite{naseer2024enabling} and complexities generated by superior IT \cite{benbya2020complexity}.  This kind of discussion is already underway in the U.S. 

These challenges must be identified and addressed in all types of complex decision-making processes, including intelligence and security. 

AI is already on its way to profoundly changing our lives, and it will do so even more in the future. Even today, advanced machine learning is an extremely powerful tool for enhancing data collection and analysis \cite{fingar2010}.  AI will enable a new dimension of comprehensive data collection and near real-time deep analysis if, but only if, adequate digitized databases have been created beforehand. The digitization and formatting of knowledge to the greatest extent possible is a prerequisite for meaningful AI support of the intelligence cycle. The recent armed conflict in Gaza has demonstrated the potential and new dimensions of AI-based/enhanced military targeting. Notwithstanding the drawbacks, we will have to acknowledge that whoever masters AI-based or AI-assisted capabilities will have a significant qualitative advantage over competitors or adversaries. They will also be able to link up with allied advanced services, others will not. 

A capability gap would risk the failure of political or military interoperability among allies when they need it most. Again, a mixed cluster of competence, staffed by ``nerds'' and professionals, will be needed to ensure workable solutions and their timely and effective implementation.

\subsection{Intelligence Service Oriented Approach (ISOA)}
The integration of AI and quantum computing into services will imply an urgent need for widespread training, awareness raising and acculturation of a workforce that will be exposed to the ``brave new world'' in its entirety, albeit with varying degrees of intensity. We must not wait until the systems are in place and ready for action, we must start much earlier, i.e. now, with a long-term perspective on the psycho-social conditions, mentalities and skills of a workforce that must be diverse, but as such will not follow the ``nerdish'' path in IT.  This was the starting point three years ago when the idea of the ICE module was born. Excellence in subject matter must be matched, supported and driven by excellence in advanced, IT-supported and IT-driven tradecraft. IT must be an integral part of staff awareness and capability building from the outset. Lack of investment and performance in the area of training jeopardizes the entire process. The curriculum is designed to trigger or encourage this kind of awareness and action orientation in its participants.

\section{ICE Module “Cyber in its Implications for Intelligence, Analysis and Decision Making”}
In line with these considerations, this cyber intelligence module is designed for middle management with responsibility for the cyber sector itself, situation analysis, crisis management or security. It is designed to provide a general, but well-founded, critical understanding of the core aspects and characteristics of the modern, globally interconnected and interdependent world in the cyber and information domain, focusing on its potential and interdependencies, as well as the resulting opportunities and risks. Within the limited framework of a five-day course (see Figure 1), the underlying goal can only be to provide food for thought and incentives for further qualification and capability building ``at home''. 

\begin{figure}[p]
\centering\includegraphics[width=\textwidth]{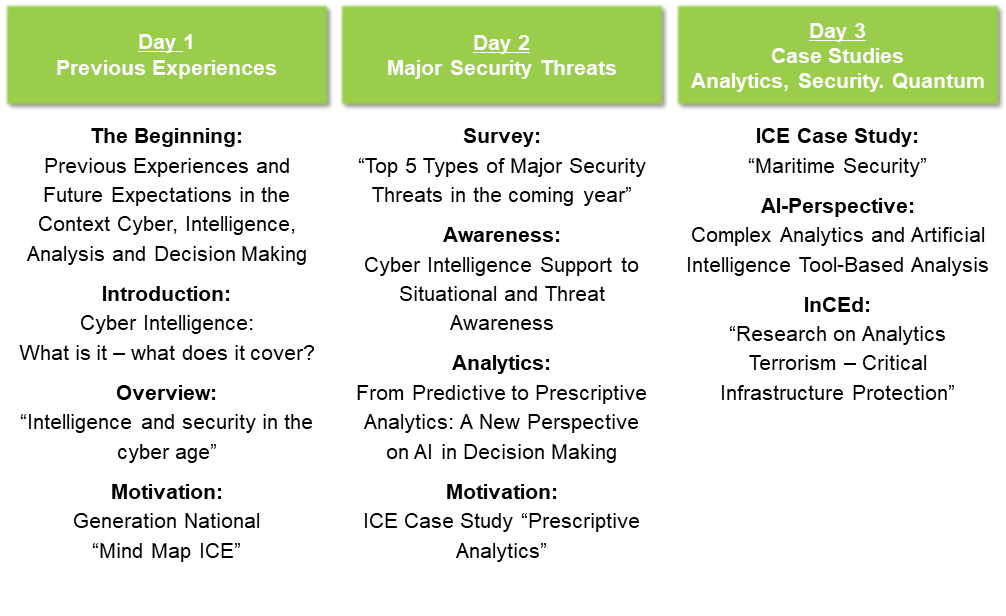}
\centering\includegraphics[width=\textwidth]{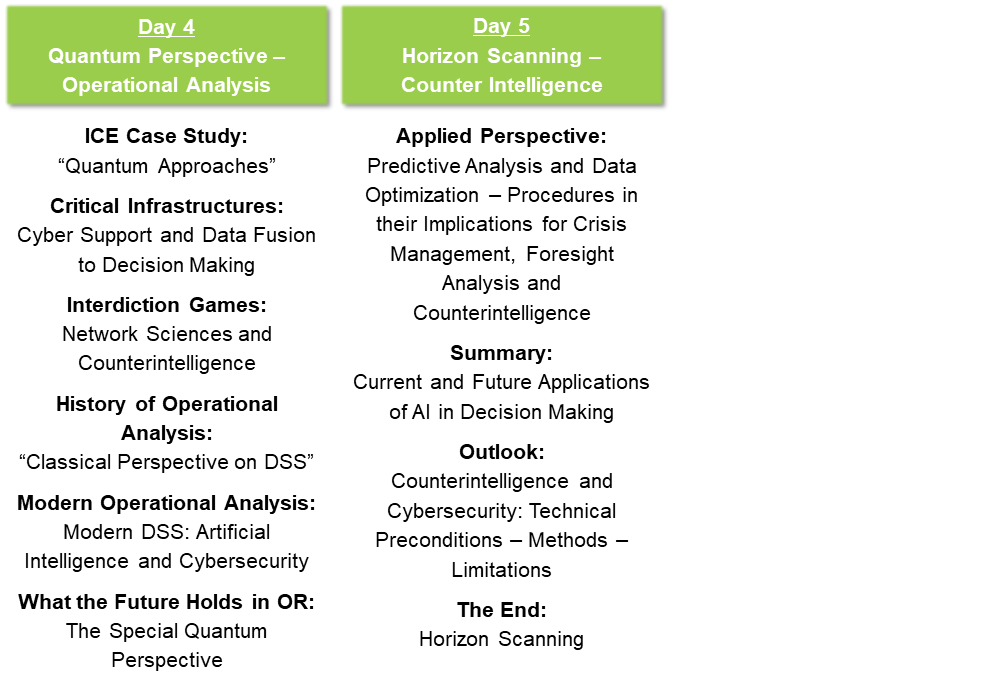}
\caption{Schedule of the five-day course “Cyber in its Implications for Intelligence, \\ Analysis and Decision Making” (CIA{\tiny D}M).}\label{CIAdM}
\end{figure}

Theoretical and practical skills are presented to identify and assess threats, vulnerabilities and risks in digitization, information processing and transmission, especially in IT infrastructures and networks. Students will become familiar with the procedures and methods of Operations Research (OR) used to protect the confidentiality, integrity and availability of general information and complex systems, including the scope of their respective performance. 

In particular, students should be able to understand and evaluate various methods and mechanisms, such as anonymization and obfuscation of communications, that make it difficult or impossible to collect information. The fundamentals of applied cryptography, data mining, system dynamics, and interdiction games will be introduced and applied to their respective areas of responsibility. 

The goal is to develop an overall innovative system-oriented understanding, to identify the essential framework conditions and influencing factors for intelligence and security services and their decision-makers in the global cyber world, and to promote the ability to model and simulate their mode of operation \cite{perko2017}.  

This interdisciplinary approach combines IT, Operations Research and Policy; it transcends conventional intelligence or cyber courses and promotes the concept of ``Support to Decision Making in the Cyber World''.

\subsection{Way-of-thinking in the Cyber World -- A Five Days Curriculum}
The course provides an interdisciplinary basic understanding of the way of thinking in cybersecurity. It covers modern analytical approaches such as data mining \cite{costalli2021},  data-driven optimization, and big data for forensic procedure development \cite{adedayo2016big},  with a special focus on their interrelationships and implications for security policy. In addition, an overview is given of current methodologies such as AI, quantum computing, and predictive analytics and optimization techniques, which are available as so-called analytical tools within ``intelligence security''. In this context, their implications for crisis management, foresight analysis, and counterintelligence will be addressed. 

The module consists of 25 sessions/seminars of two hours each. Depending on prior agreement with the interested services, a further 75 hours of private study could be devoted to the preparation of a short individual presentation and a written paper of 2,500 words on a topic to be defined in advance with the module convenor. One or two (international) excursions or active business game phases (wargames) are planned as part of the program.


\subsection{Key findings and lessons learned from the pilot course}
After two rounds of implementation, the evaluations confirm the comments and recommendations made earlier. Participants have learned that cyber activities are not only a technical matter, but also have an important socio-cultural context. Therefore, social and regional sciences will have always to be included in the assessment of cyber and its relevant actors. This became particularly clear in the embedded case studies. A major challenge remains to combine a situation picture based on structured data with information from unstructured data \cite{BorghoffS96}. The following questions were intensively discussed by the participants:
\begin{itemize}
\item
How do you transform unstructured data into structured data without losing or altering the content? 
\item
How can this process be accelerated using OR-techniques to support data-driven decision making? 
\item
In particular, could AI powered by quantum computing be a way forward?
\end{itemize}

\subsection{Optimizing situational awareness -- An OR-based threat analysis}
Situational awareness may need to be organized in a structured bottom-up manner, starting at the regional and sectoral (civil, military, police, critical infrastructure) levels and ending at the federal/national government level. 

\begin{figure}[ht]
\centering\includegraphics[width=\textwidth]{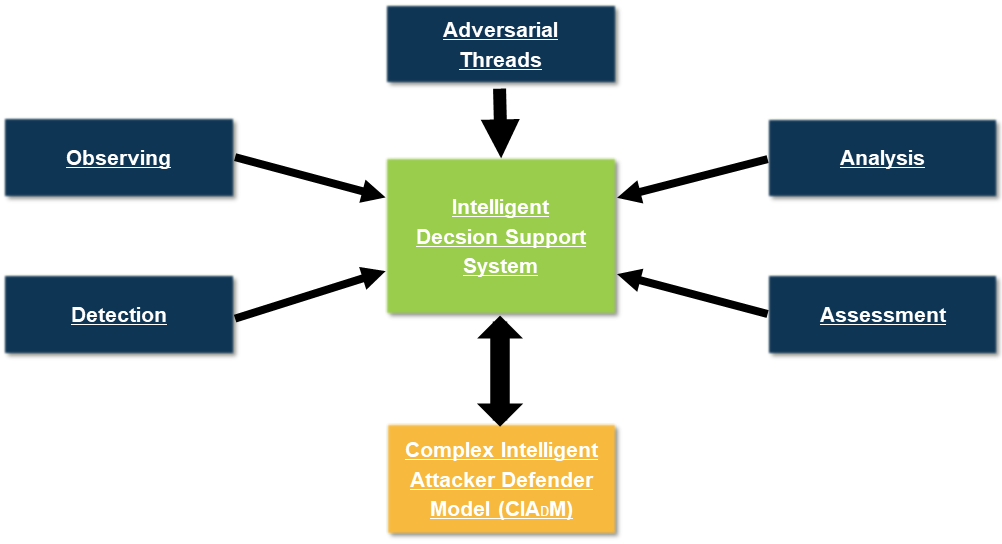}
\caption{Intelligent decision support system.}\label{Folie3}
\end{figure}

Strong cyber connectivity and intergovernmental upstream awareness would be a prerequisite for this comprehensive support to coordinated multi-agency decision making. As shown in Figure~\ref{Folie3}, cyber support for OR-based decision-making processes by providing comprehensive and timely analyzed data is increasingly critical at all stages of the trained and educated process. 

Tactical early warning of potentially imminent threats must be based on a comprehensive set of quantitative and qualitative hard intelligence data on adversary capabilities (military or hybrid tools) in an objective threat configuration. This requires long-term, all-source intelligence collection and analysis in advance, and databases that are accessible in real time to contribute to the required situational awareness. Data analysis, including data farming, plays a critical role in this process. The same is true for establishing and assessing situational awareness of one's own capabilities and vulnerabilities (attacker-defender models). 

Strategic early warning must be based on comprehensive data sets that correlate with the potential threat/risk. The selection of these data for data mining or data farming processes is of paramount importance for the timeliness and pertinence of a result. Data science combined with subject matter expertise should result in meaningful and actionable indicators. 

\subsection{Center of Excellence between Control Towers and Living Labs}
Indicator-based early warning functions and services have been developed in the EU (EU Early Warning System \cite{joint2021} for crisis and conflict \cite{fact2020}) and in Germany (for example: Preview, COMTESSA\footnote{https://www.unibw.de/comtessa}  Suite II-CE-LLab: Intelligent Indicator-based Control Experimental Living Lab, “REVEARS”). Data Analytics \cite{carrascosa2017data} (if possible AI-supported, combined with machine learning, reinforcement learning, game theory) is essential in defensive cybersecurity (detection of irregularities in one's own system) and in proactive cyberdefense (detection of intrusion sets where they occur (e.g. through SIGINT support to cyberdefense) \cite{janeja2022data}.  In addition, data integrity is becoming increasingly critical to operational security and outcomes in the face of growing cyber-enabled capabilities to corrupt data content. This is all the more important in the case of cyber-enabled situational awareness that needs to reach back to databases in real time. Major efforts are needed to preserve and protect data integrity, including the provision of strong and independent backbone capabilities.

Quantum computing will be a game changer when it is operational on a larger scale. Large-scale hardware and software development will take decades. Broad-based hardware procurement would be premature given the dynamics of innovation and paradigm shifts. However, the recruitment and development of highly qualified experts capable of participating in the dynamics of the development of reinforcement and quantum computing must begin now at the latest. 

The creation of specialized centers of excellence (services/science) and further specialized curricula/training modules would be an important step. Judgement-based analysis embedded in the constitution of control towers and living labs for data-driven decision support systems is central.

\subsection{Outlook: How to get started?}
The imperative of transformation to cope with the age of digitalization is the starting point and basis of this contribution.

Transformation is the responsibility of political and service leadership, from start to finish. It is a delicate process with multiple risks of hick-ups and failures that could affect the core functionality of the services in providing situational awareness and security. Transformation must not be a botched process that leaves the organization, or parts of it, in shambles. Therefore, some requirements for initiating a successful process are: 
\begin{itemize}
\item
Choosing the right leader with vision, a strong sense of realism and the willingness and ability to live with the results created by the process.
\item
Creating a core group of like-minded top management. 
\item
Creating a small but viable subordinate core staff with defined responsibilities for a coherent set of action items. 
\item
Begin to involve staff more broadly in education, training and skill building. Preferred starting points could be workshops and seminars (e.g., the proposed cyber intelligence module) for those interested, then let them spread the gospel and act as role models, supporters and advisors to their colleagues. 
\item
Closely follow, monitor and support the process. Transformation must be the priority of the boss and his/her closest management. 
\item
Education of minds and souls must be a top priority from the earliest and an everlasting exercise throughout the process and beyond.
\end{itemize}

The curriculum is designed to help make this process more sustainable, particularly by raising managers' awareness and sensitivity to the factual and managerial challenges of transformation.

\subsection{Embedding in Innovative Leadership}
Structures need to be run by a sufficient number of well-trained staff. Lean structures cannot compensate for a fundamental lack of staff quantity, and even less for a structural lack of quality and expertise. The leaner the structure, the more it demands of the staff, their professional qualifications, expertise, skills, sense of purpose and dedication. 

The essential factor in this area is competent leadership that fosters social competence and unity of purpose. Leadership is a core function in all organizations. It must be inspired and trained to provide top-down input and receive bottom-up feedback. This qualification and sense of purpose must start with the core group of top management, then be transferred by them to their respective staff/middle management and cascade down to the working levels. 

The essential core is to achieve unanimity in aims and values. This process must result in a unified and coherent set of aims and a strong sense of shared responsibility. In addition, internal leadership has to be supported from the outside, by the superior political authorities in government and parliament, but also by public opinion. The cyber intelligence module reflects these needs and is designed to contribute to such an innovative leadership approach in a holistic way.

\section{Conclusion}
We present the concept of the ICE and its role in fostering cooperation among European national intelligence services to address today's security challenges. ICE has developed a comprehensive academic program aimed at training intelligence professionals and promoting interdisciplinary dialogue. It emphasizes the importance of adapting to rapid technological advances, particularly in the areas of cyber intelligence and AI. We emphasize the need for continuous learning and adaptation to evolving threats. We delve into the challenges and opportunities presented by digitalization, AI, and quantum computing in the realm of intelligence collection and decision-making. We highlight the need for intelligence organizations to enhance their capabilities in data analysis, cybersecurity, and predictive analytics. The integration of AI and quantum computing is seen as critical to maintaining information superiority and optimizing decision making.

The ICE module on ``Cyber in its implications for Intelligence, Analysis and Decision Making'' is outlined, focusing on developing skills in threat assessment, data analysis, and decision support. The curriculum adopts an interdisciplinary approach, combining IT, operations research, and policy perspectives. Key findings from the pilot course underscore the importance of integrating socio-cultural factors into cyber activities and the challenges of transforming unstructured data into actionable intelligence. Finally, we discuss the imperative of transformation in intelligence organizations and the role of leadership in driving change. We advocate for a collaborative and innovative leadership approach that fosters unity of purpose and social competence. The proposed curriculum is positioned to support this transformative process by raising awareness and providing practical tools for effective leadership in the digital age.

\section*{About the authors}
{\bf Lars Berger} is professor at the Department of Intelligence at the Federal University of Administrative Sciences in Berlin, academic adviser to the Intelligence College in Europe (ICE) as well as Germany’s academic representative within the ICE. 
The author can be contacted at lars.berger@hsbund-nd.de.
\newline
https://orcid.org/0000-0001-7846-804X
\medskip

\noindent
{\bf Uwe M. Borghoff} is professor at the Department of Computer Science at the University of the Bundeswehr Munich. From 2018 to 2024, he was the founding director of the Center for Intelligence and Security Studies (CISS Munich) and course director of the Master in Intelligence and Security Studies (MISS). 
The author can be contacted at uwe.borghoff@unibw.de.
\newline
https://orcid.org/0000-0002-7688-2367
\medskip

\noindent
{\bf Gerhard Conrad} is professor at the Department of Intelligence at the Federal University of Administrative Sciences in Berlin. From 1990 to 2019, he was a civil servant in the German Federal Intelligence Service (BND), where he performed a variety of tasks in analysis and operations in the Near/Middle East, including as a mediator in humanitarian affairs between Israel, Hizballah, and also Hamas.
The author can be contacted at gerhard.conrad@hsbund-nd.de.
\newline
https://orcid.org/0009-0001-3770-8032
\medskip

\noindent
{\bf Stefan Pickl} is professor at the Department of Computer Science at the University of the Bundeswehr Munich. Since 2019, he has headed the cyber intelligence module in MISS as well as in the academic program of ICE.
The author can be contacted at stefan.pickl@unibw.de.
\newline
https://orcid.org/0000-0001-5549-6259

%

%
%
%

\end{document}